\documentclass[twocolumn,aps,superscriptaddress,nofootinbib]{revtex4}
\usepackage[dvips]{graphicx}
\usepackage{latexsym,amssymb,amsmath}
\usepackage{color}
\usepackage{enumerate}
\usepackage[bookmarksnumbered,bookmarksopen,colorlinks,citecolor=red,linkcolor=blue]{hyperref}
\usepackage{mathrsfs}
\usepackage{times}
\usepackage{csquotes}
\usepackage{multirow}
\usepackage{diagbox}
\usepackage{booktabs}
\usepackage[dvipsnames]{xcolor}

\graphicspath{{./figs/}}

\begin{document}

\title{Bottom energy loss and non-prompt $J/\psi$ production in relativistic heavy ion collisions}

\author{Meimei Yang}
\affiliation{Department of Physics, Tianjin University, Tianjin 300354, China}

\author{Shiqi Zheng}
\affiliation{University of North Carolina at Chapel Hill, North Carolina 27599, USA}

\author{Bo Tong}
\affiliation{Department of Physics, Tianjin University, Tianjin 300354, China}

\author{Jiaxing Zhao}\email{zhao-jx15@tsinghua.org.cn}
\affiliation{SUBATECH, Universit\'e de Nantes, IMT Atlantique, IN2P3/CNRS, 4 rue Alfred Kastler, 44307 Nantes cedex 3, France}

\author{Wenyuan Ouyang}\email{oywy@tju.edu.cn}
\affiliation{Department of Physics, Tianjin University, Tianjin 300354, China}

\author{Kai Zhou}\email{zhou@fias.uni-frankfurt.de}
\affiliation{Frankfurt Institute for Advanced Studies, Giersch Science Center, D-60438 Frankfurt am Main, Germany}

\author{Baoyi Chen}\email{baoyi.chen@tju.edu.cn}
\affiliation{Department of Physics, Tianjin University, Tianjin 300354, China}

\date{\today}
\begin{abstract}
We study the momentum and centrality dependence of the non-prompt 
$J/\psi$ nuclear modification factors ($R_{AA}$), which comes from the $B$ hadrons decay, in Pb-Pb collisions 
at the Large Hadron Collider.  Bottom quarks are produced in the parton hard scatterings and suffer energy loss in the quark-gluon plasma and the hadronic gas, where the spatial and time evolution of the medium 
is described with the hydrodynamic equations. Medium-induced 
elastic scatterings and the radiation in bottom quarks  
are included in the energy loss of bottom quarks. The hadronization process of bottom 
quarks is described with the instantaneous coalescence model. After considering both cold and hot nuclear matter effects in Pb-Pb collisions at $\sqrt{s_{NN}}=5.02$ TeV, 
we calculated the $R_{AA}$ and also the elliptic flows of non-prompt $J/\psi$ from the decay of 
$B$ mesons at different centralities and transverse momentum bins. The $R_{AA}$ and $v_2$ of non-prompt $J/\psi$ sensitive to the hot medium reflect and centrality supply an opportunity to study the bottom quarks energy loss in the hot medium.
\end{abstract}
\maketitle

\section{Introduction}
\label{sec:introduction}

A new state of matter consisting of quarks and gluons called Quark-Gluon Plasma (QGP), 
is believed to be produced in the nucleus-nucleus collisions~\cite{Bazavov:2011nk} performed at the Relativistic Heavy-Ion Collider (RHIC) and the Large Hadron Collider (LHC). 
Studying the properties of the deconfined medium 
helps to understand the Quantum Chromodynamics (QCD) at finite temperatures~\cite{Satz:2000bn,Bass:1998vz,Cassing:1999es,Shuryak:1983ni}. 
Heavy quarks and quarkonium which are produced in the 
initial parton hard scatterings are sensitive to not only properties of the QGP~\cite{Matsui:1986dk,Zhao:2010nk,Du:2015wha,Liu:2010ej,Zhou:2014kka,Chen:2018kfo,Zhao:2017yan,Yao:2020eqy,Blaizot:2015hya,Brambilla:2020qwo,Delorme:2022hoo,Miura:2022arv} but also the early stage of heavy-ion collisions~\cite{Zha:2018ytv,Shi:2017qep,Zhao:2021voa,Chen:2019qzx}, see recent review papers~\cite{Rothkopf:2019ipj,Zhao:2020jqu}.
Heavy quarks dump energy to the medium via radiation and the 
random elastic scatterings with 
thermal partons when 
they move inside the QGP~\cite{Djordjevic:2003zk,Wicks:2005gt,Braaten:1991we,Moore:2004tg,Gyulassy:2000fs,Arnold:2002zm,Zhang:2003wk,Qin:2007rn,Majumder:2009ge,Berrehrah:2014kba}. The magnitude of the medium-induced energy loss depends 
on the coupling strength between heavy quarks and the medium and also the initial medium energy density, which are all encoded in the heavy quark transport coefficients~\cite{Cao:2018ews,Rapp:2018qla,Gossiaux:2010yx,He:2012df,Das:2012ck,Mazumder:2013oaa,Song:2019cqz,Kurian:2020orp,Prakash:2021lwt}. 
When the local temperature of the medium drops to a critical 
value, heavy quarks hadronize into mesons or baryons via the recombination with light quarks or the fragmentation~\cite{Cao:2013ita,He:2019vgs,Gossiaux:2009mk,Plumari:2017ntm}. In the hadronic phase, open heavy flavor hadrons continue losing energy via collisions with thermal light hadrons~\cite{Fuchs:2004fh,He:2011yi}. Therefore, 
the nuclear modification factors of $B$ and $D$ mesons have been used to extract the transport 
coefficients in phenomenological studies~\cite{Scardina:2017ipo,Cao:2018ews,Dong:2019unq}. 

In experiments, the nuclear modification factor and the elliptic flow of $D$ mesons have been measured~\cite{ALICE:2013olq,ALICE:2020iug,CMS:2020bnz}, which indicates a strong coupling between heavy quarks and the QGP, which created in the heavy ion collisions at the LHC. 
Except for the $b$-decayed electrons or muons, there is no direct observation of the nuclear modification factor and the elliptic flow of $B$ mesons in heavy ion collisions. However, the non-prompt $J/\psi$ from the decay of $B$ hadrons has been measured by the CMS and ATLAS Collaboration~\cite{CMS:2017uuv,ATLAS:2018xms}. This supplies a new way to study bottom quark energy loss and hadronization in the QGP. 
Due to the energy loss, the high transverse momentum bottom quark will shift to the low transverse momentum region.
This will suppress the final spectrum of bottom quarks at high $p_T$ while enhancing the yield in the low $p_T$. At LHC energies, there is a sizable modification in the parton distribution function in the nucleus~\cite{Eskola:2009uj}, which affects the production of bottom quarks directly. This is called the shadowing effect and is cast into the cold nuclear matter effect. It suppresses the initial production of bottom quarks and changes the nuclear modification factors of $B$ hadrons or non-prompt $J/\psi$. 

In this work, we take the momentum distribution of the initial distribution of bottom quarks given by the Fixed-Order-Next-to-Leading-Log (FONLL) calculation~\cite{Cacciari:1998it,Cacciari:2012ny}. 
The cold nuclear matter effect is considered by modifying the initial distribution before the start of Langevin dynamics for heavy quark diffusion in the QGP. The parameters in the Langevin equation such as the transport coefficients are determined in previous works~\cite{Chen:2017duy}. When the local temperature of the bottom quark is smaller than the critical temperature, the bottom quark will hadronize into $B$ hadrons. These $B$ hadrons continue to lose energy in the hadronic phase and finally decay into the non-prompt $J/\psi$ after the kinetic freeze-out.

This paper is organized as follows. 
In Sec.~\ref{sec.framework}, we will describe the theoretical framework of heavy quark evolution and hadronization. In Sec.~\ref{sec.hydro}, hydrodynamic equations are presented to simulate the 
realistic expansion of the hot medium created in heavy-ion collisions. This is treated as a background medium of the heavy 
quark diffusion. Numerical results about the nuclear modification factor $R_{AA}$ and elliptic flow $v_2$ of non-prompt $J/\psi$ in different centralities and transverse momentum are calculated and shown in Sec~\ref{sec.results}. We summarize in Sec.~\ref{sec.summary}. 

\section{Dynamical evolutions of bottom quarks and hadronization}
\label{sec.framework}

In the QGP, bottom quarks will scatter with thermal partons. With a small momentum transfer in each collision, the dynamical evolution of bottom quarks can be treated as Brownian 
motion. The stochastic evolution of heavy quark momentum 
have been well described by the Langevin equation~\cite{He:2012df,Cao:2013ita,Das:2015ana,Chen:2017duy}. The medium-induced gluon radiation is also included in the energy loss of heavy quarks, which becomes dominant for heavy quarks with large momentum. Correspondingly the Langevin equation for bottom quark dynamics can be written as~\cite{Cao:2013ita,Cao:2015hia},
 \begin{align}
\label{lan-gluon}
{d{\bf p}\over dt}= -\eta(p) {\bf p} +{\bf \xi} + {\bf f}_g,
\end{align}
with $\bf p$ is the bottom quark momentum. 
The drag term $\eta(p)$
is connected with the momentum diffusion coefficient via 
the fluctuation-dissipation relation, 
$\eta(p)=\kappa/(2TE_b)$, where the bottom quark energy 
is $E_b=\sqrt{m_b^2+{\bf p}^2}$ and {$m_b=4.5$ GeV} is the bottom quark mass. 
The momentum diffusion coefficient $\kappa$ is related to the spatial diffusion coefficient $\mathcal{D}_s$ through, $\kappa \mathcal{D}_s = 2T^2$. In the QGP and the 
hadronic medium, the spatial diffusion coefficient 
is used as $\mathcal{D}_s(2\pi T)=7$ and $9$ in this study respectively~\cite{Dong:2019unq}. 
The stochastic term ${\bf \xi}$ is treated as white noise. Neglect the 
momentum dependence in the ${\bf \xi}$, it satisfies the 
relation, 
\begin{align}
\langle \xi^{i}(t)\xi^{j}(t^\prime)\rangle =\kappa \delta ^{ij}\delta(t-t^\prime), 
\end{align}
where the index $i,j$ represents three dimensions. 
The noise terms 
are uncorrelated at different time points. Gluon radiation contribution is represented by ${\bf f}_g=-d{\bf p}_g/dt$ with ${\bf p}_g$ the momentum of the emitted gluon. 
The number of emitted gluons in the time interval $t\sim t+\Delta t$ is~\cite{Cao:2013ita}, 
\begin{align}
\label{gluon-spec}
P_{\rm rad}(t,\Delta t) = \langle N_g(t, \Delta t)\rangle = \Delta t \int dx d k_T^2
{dN_g\over dx dk_T^2dt}.
\end{align}
In the numerical simulation with small $\Delta t$, the number of the emitted gluon in the period 
is smaller than the unit and $P_{\rm rad}$ can be 
treated as the 
radiation probability. 
The variable $x=E_g/E_b$ is the ratio of the energies of 
the emitted 
gluon and the bottom quark.
$dN_g/dxdk_T^2dt$ is the 
the spectrum of emitted gluon given by perturbative QCD calculations~\cite{Zhang:2003wk,Majumder:2009ge}. $k_T$ is the transverse momentum 
of the gluon. At each time step, the bottom quark momentum is updated with the Langevin equation.  

The initial momentum 
distribution of bottom quarks in heavy ion 
collisions can be simulated via the FONLL model~\cite{Cacciari:1998it,Cacciari:2012ny}.
For the initial spatial distribution of bottom quarks, as they are produced in the parton hard scatterings, the initial spatial density of bottom quarks is proportional to the 
product of two thickness function $T_A({\bf x}_T+{\bf b}/2)T_B({\bf x}_T-{\bf b}/2)$, where ${\bf b}$ is the impact 
parameter. 
In heavy-ion collisions, the parton density in the nucleus 
is changed relative to the distribution in the free nucleon, where 
the bottom yield is also changed. The ratio 
of the parton density in the nucleus $f_A$ and the free nucleon 
$f_n$ is defined as $\mathcal{R}=f_A(x, \mu_F)/(Af_n(x,\mu_F))$. The longitudinal momentum fractions of the two initial gluons are defined as $x_{1,2}=e^{\pm y}\sqrt{m_{b\bar b}^2+(2p_T)^2}/\sqrt{s_{NN}}$ in nucleus A and B respectively. 
$p_T$ and $y$ are the transverse momentum and the rapidity of one bottom quark. $m_{b\bar b}$ is the total mass of bottom and anti-bottom quarks. 
$\mu_F=\sqrt{m_{b\bar b}^2+(2p_T)^2}$ is the factorization scale. 
The shadowing modification factor of bottom quarks is calculated with the 
EPS09 package~\cite{Eskola:2009uj} in this work. The global shadowing factor of 
bottom quarks is 
calculated to be $0.86\sim 0.95$ in the nucleus. The spatial 
dependence of the shadowing factor is included which is 
proportional to the nuclear thickness function $T_{A(B)}({\bf x}_T)$~\cite{Zhou:2014kka}, which are calculated from optical 
Glauber model~\cite{Miller:2007ri}. The initial distribution of 
bottom quarks in the Pb-Pb collisions with the modification 
of the shadowing effect 
is written as 
\begin{align}
{d^3N^{b\bar b}_{AA}\over d{\bf x}_Tdyd{ p}_T}=&{d^2\sigma^{b\bar b}\over dydp_T}
T_A({\bf x}_T+{\bf b}/2)T_B({\bf x}_T-{\bf b}/2)\nonumber \\
&\times\mathcal{R}_A({\bf x}_T,x_1,\mu_F)\mathcal{R}_B({\bf x}_T,x_2,\mu_F).
\label{lab-AA-init}
\end{align}
In the event-by-event numerical simulations, we randomly generate 
a large set of bottom quarks based on the Eq.(\ref{lab-AA-init}) 
with different positions and momentum. Each bottom quark is 
evolved under the Langevin equation Eq.(\ref{lan-gluon}).

When the bottom quark travel across the hadronization boundary, which is defined by the local temperature equaling the critical temperature, $T=T_c$, the bottom quark will convert to $B$ hadrons. The hadronization mechanism used in this study is the coalescence model.
The momentum of bottom quarks at the hadronization
is determined by the Langevin equation. The 
the momentum of thermal light quarks used in the coalescence process is generated with the Fermi distribution. 
The coalescence probability of bottom quarks and light quarks to form $B$ mesons can be described with~\cite{Chen:2021akx}, 
\begin{align}
\label{eq-Dcoal}
&\mathcal{P}_{b+\bar q\rightarrow B}({\bf p}_B)\nonumber \\
&=\int {d{\bf p}_1\over (2\pi)^3} {d{\bf p}_2\over (2\pi)^3}
{dN_1\over d{\bf p}_1} {dN_2\over d{\bf p}_2 }
f_B^W({\bf q}_r)\delta^{(3)}({\bf p}_B -{\bf p}_1-{\bf p}_2), 
\end{align}
where $dN_1/{d\bf p}_1$ is the normalized momentum distribution of 
bottom quarks at the hadronization hypersurface. $dN_2/d{\bf p}_2$ is the normalized Fermi distribution of light quarks 
in the local rest frame (LRF) of the expanding medium. It 
is $f({\bf p}_2^{\rm lrf})
=N_0/(e^{\sqrt{(m_2^2+|{\bf p_2^{\rm lrf} } |)/T_c}} +1)$ with $N_0$ is 
the normalization factor. The thermal mass of light quarks 
is set to be $m_2=0.3$ GeV~\cite{Chen:2021akx}. In this 
setup, $\mathcal{P}_{b+\bar q\rightarrow B}({\bf p}_B)$ 
is understood as the coalescence probability of one bottom 
quarks with the momentum ${\bf p}_1$ 
turning into a $B$ meson with the momentum ${\bf p}_B$.  
$\delta$-function ensures the momentum conservation in the 
reaction, 
where the momentum of emitted gluon has been 
neglected to get the simplified relation ${\bf p}_B= {\bf p}_1+{\bf p}_2$. The Wigner function $f_B^W({\bf q}_r)$ depends on the 
relative momentum between the bottom and the thermal 
light quark. 
In the center-of-mass (COM) frame of the formed $B$ meson, {
${\bf q}_r=(E_2^{\rm cm}{\bf p}_1^{\rm cm}- E_1^{\rm cm}{\bf p}_2^{\rm cm})/(E_1^{\rm cm}+E_2^{\rm cm})$}. ${\bf p}_1^{\rm cm}$ and ${\bf p}_2^{\rm cm}$ are the momentum of the bottom quark and the light quark 
in the COM frame. 
$E_1^{\rm cm}$ and  $E_2^{\rm cm}$ are the corresponding energies. 
The Wigner function is taken as a Gaussian function with the width 
determined by the mean square radius of the $B$ meson,
$\sigma^2= {4\over 3}{(m_1+m_2)^2\over m_1^2 +m_2^2}
\langle r^2\rangle_B $~\cite{Song:2016lfv}.
We approximate the root-mean-square radius of $B$ meson as
$\sqrt{\langle r^2\rangle_B}=0.43$ fm. $B$ meson production is dominated by the coalescence process at low $p_T$ regions.  From model calculations~\cite{Cao:2015hia} about bottom quark hadronization, the coalescence probability is much larger than the fragmentation in the low region, such as $p_T\lesssim 5$ GeV/c. {In this work, we take a simplification and assume that all $B$ mesons are produced via the coalescence process in this low $p_T$ region. While at higher $p_T$, all $B$ mesons are produced via the fragmentation where the momentum of B meson is assumed to be the momentum of bottom quark.}

After the hadronization, the $B$ meson will evolve in the hadronic phase and finally decay weakly into various hadrons. Many of the decay channels are difficult to be observed in the heavy ion collisions except the $B\to J/\psi$. This component is named non-prompt $J/\psi$, which makes a significant contribution to the inclusive $J/\psi$, especially at high $p_T$ regions~\cite{CMS:2017uuv,ATLAS:2018xms}. And this offers the opportunity to study the bottom quark energy loss and hadronization in the QGP. In the real case, the bottom quark will convert to various $B$ hadrons, such as $B^{0,\pm}$, $B_s$, $\Lambda_b$, and so on. Based on the thermal model~\cite{Andronic:2021erx}, a simple estimation shows there are more than 60\% bottom quarks become $B$ mesons, including $B^{0,\pm}$ and $B_s$. The observed data shows the branching ratio of $B$ meson to $J/\psi$ is larger than bottom baryons~\cite{10.1093/ptep/ptaa104}. So, as a first approximation, we force all bottom quarks to convert to $B$ mesons and multiply a 60\% factor to the final $B$ meson spectra in this study.
\begin{figure}[!hbt]
\centering
\includegraphics[width=0.38\textwidth]{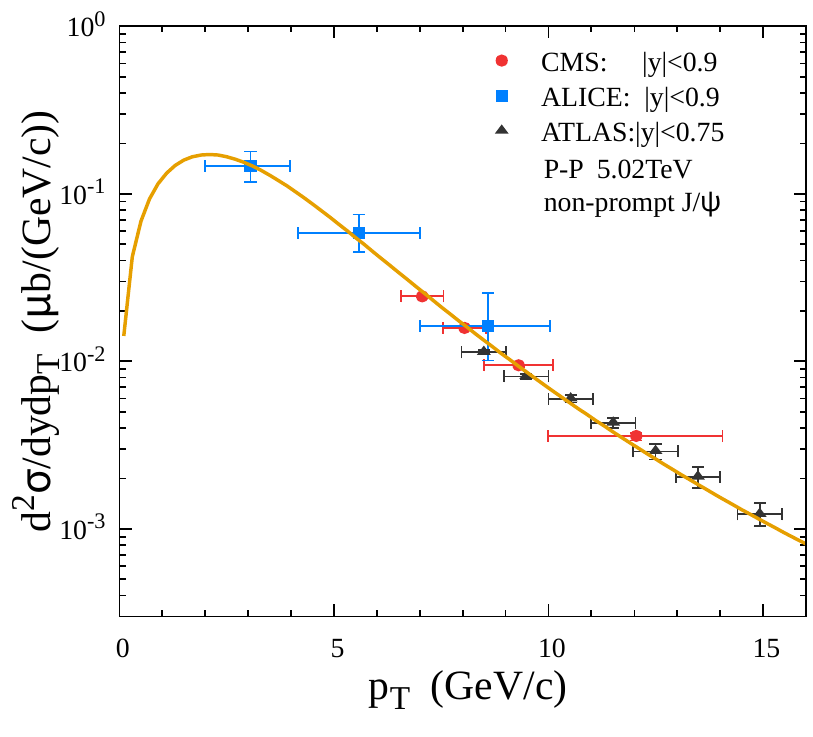}
\caption{ The transverse momentum spectrum $d^2\sigma/dydp_T$ of non-prompt 
$J/\psi$ in the central rapidity of proton-proton collisions 
at $\sqrt{s_{NN}}=5.02$ TeV. The experimental data are 
cited from the ALICE~\cite{ALICE:2021edd}, CMS~\cite{CMS:2017exb}, and ATLAS~\cite{ATLAS:2017prf} Collaborations.
}
\label{fig-npjpsi}
\end{figure}

The unknown thing is the decay branch fraction $\mathcal{F}$ of $B$ mesons to $J/\psi$. We estimate this fraction in A-A collisions is the same as the p-p collisions. So, it can be extracted from p-p data,
\begin{align}
 \mathcal{F}(p_T)={dN^{\rm nonprompt}_{J/\psi}/dp_T \over dN_B/dp_T},
\label{lab-fraction}
\end{align}
where the $B$ meson spectra $dN_B/dp_T$ in p-p collisions are observed by the ALICE~\cite{ATLAS:2013cia} and CMS~\cite{CMS:2011oft} Collaborations.
The differential cross section of non-prompt $J/\psi$ in 5.02 TeV pp collisions 
have been measured by the ALICE~\cite{ALICE:2021edd}, CMS~\cite{CMS:2017exb}, and ATLAS~\cite{ATLAS:2017prf} Collaborations. The transverse momentum distribution can be fitted with the below formula~\cite{Chen:2018kfo},
\begin{align}
    {d^2\sigma\over dy dp_T}&={dN^{\rm norm}\over dp_T}\cdot {d\sigma\over dy},\nonumber \\
    {dN^{\rm norm}\over dp_T}&= {2\pi p_T (n-1)\over \pi (n-2) \langle p_T^2\rangle_{pp}} [1+{p_T^2\over (n-2)\langle p_T^2\rangle_{pp}}]^{-n},
    \label{eq-init-pt}
\end{align}
where $dN^{\rm norm}/dp_T$ is the normalized transverse momentum 
distribution. Rapidity differential cross section $d\sigma/dy$ will be eliminated in the calculation of $R_{AA}$. The 
parameters $n$ and $\langle p_T^2\rangle_{pp}$ controls the 
the shape of the line. They are determined as 
$n=3.2$ and $\langle p_T^2\rangle_{pp}=19.5\ \rm{(GeV/c)^2}$. The data and fitted curve are shown in Fig.~\ref{fig-npjpsi}. The yield of non-prompt $J/\psi$ in heavy ion collisions can be obtained by multiplying the branch fraction $\mathcal{F}$ to the $B$ mesons spectra after the evolution. 

\section{Hot medium evolution}
\label{sec.hydro}
The deconfined medium produced in heavy-ion collisions 
turns out to be a strongly coupled matter. Its dynamical 
evolution has been extensively studied with hydrodynamic 
equations~\cite{Schenke:2010rr,Shen:2014vra,Pang:2012he}. We employ the 2+1 dimensional ideal 
hydrodynamic equations to describe the medium expansion 
on the transverse plane. Along the longitudinal direction 
defined as the direction of nuclear acceleration, there 
is a flat pattern in the initial rapidity distribution of the  
charged hadron multiplicities~\cite{Schenke:2010rr}. With this feature, 
the longitudinal evolution of the hot medium is 
treated to be a Bjorken expansion. The hydrodynamic equation 
is written as 
\begin{align}
\partial_{\mu\nu} T^{\mu\nu}=0,
\end{align}
where the energy-momentum tensor is 
$T^{\mu\nu}=(e+p)u^\mu u^\nu -g^{\mu\nu}p$. $u^\mu$ is 
the four-velocity of the fluid. $e$ and $p$ are the 
energy density and the pressure of the medium. 
The equation of state (EoS) of the deconfined matter is 
taken to be the EoS of the ideal gas. While the EoS 
of the hadronic 
matter is taken from 
the Hadron Resonance Gas model~\cite{Sollfrank:1996hd}. The hadronization phase 
transition is a first-order, with the critical temperature 
to be $T_c=165$ MeV. This value is determined by selecting 
the mean field repulsion parameter and the bag parameter to be 
$K=450\ \rm{MeV\ fm^3}$ and $B^{1/4}=236\ \rm{MeV}$ at the zero baryon number density~\cite{Sollfrank:1996hd}. 

\begin{figure}[!hbt]
\centering
\includegraphics[width=0.38\textwidth]{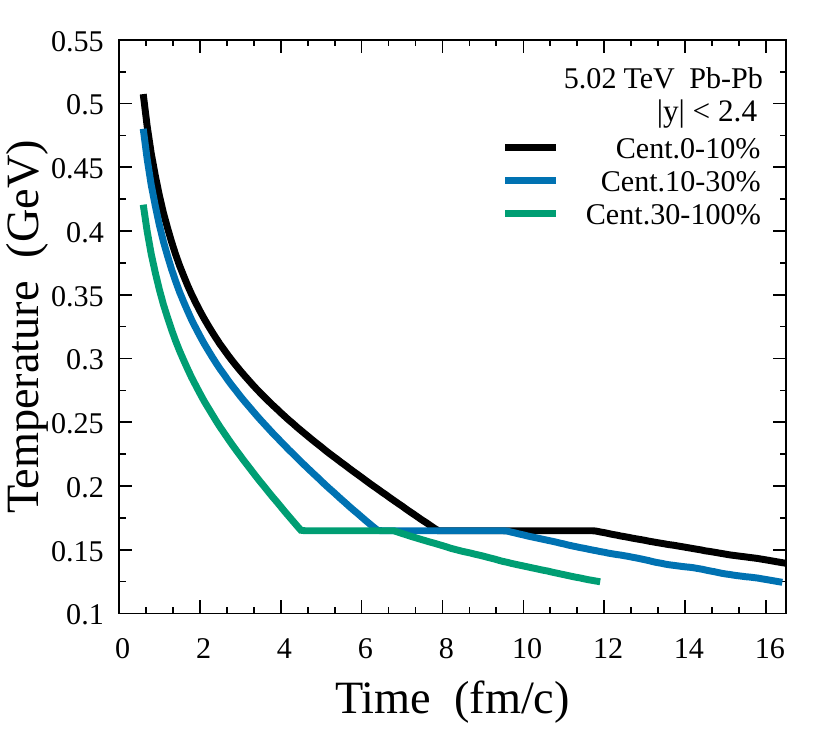}
\caption{ (Color online) The time evolution of the temperature at the center of the medium (${\bf x}_T=0$) in Pb-Pb collisions at $\sqrt{s_{NN}}=5.02$ TeV. Temperatures in different collision centralities are plotted with different color lines. The critical temperature of the deconfined phase 
transition is $T_c=165$ MeV. 
}
\label{lab-hydro}
\end{figure}
The initial  
entropy density comes from both soft and hard processes. The profile 
of the entropy density is assumed to be proportional to 
the number of nucleon participants $n_{part}({\bf x}_T)$ and the number of binary collisions $n_{coll}({\bf x}_T)$.  The 
entropy density is determined by the charged hadron multiplicities, where the maximum temperature at the center of the medium is $T_0({\bf x_T}=0,\tau_0)=510$ MeV~\cite{Chen:2018kfo} at the time $\tau_0$. 
$\tau_0$ is the time scale of the medium reaching local equilibrium. 
Its value is fitted as 
$\tau_0=0.6$ fm/c by the collective flows of light hadrons observed in Pb-Pb collisions~\cite{Shen:2014vra}. The time evolutions of the medium temperature in different centralities are plotted 
in Fig.~\ref{lab-hydro}.

\section{results}
\label{sec.results}

\begin{figure}[!hbt]
\centering
\includegraphics[width=0.38\textwidth]{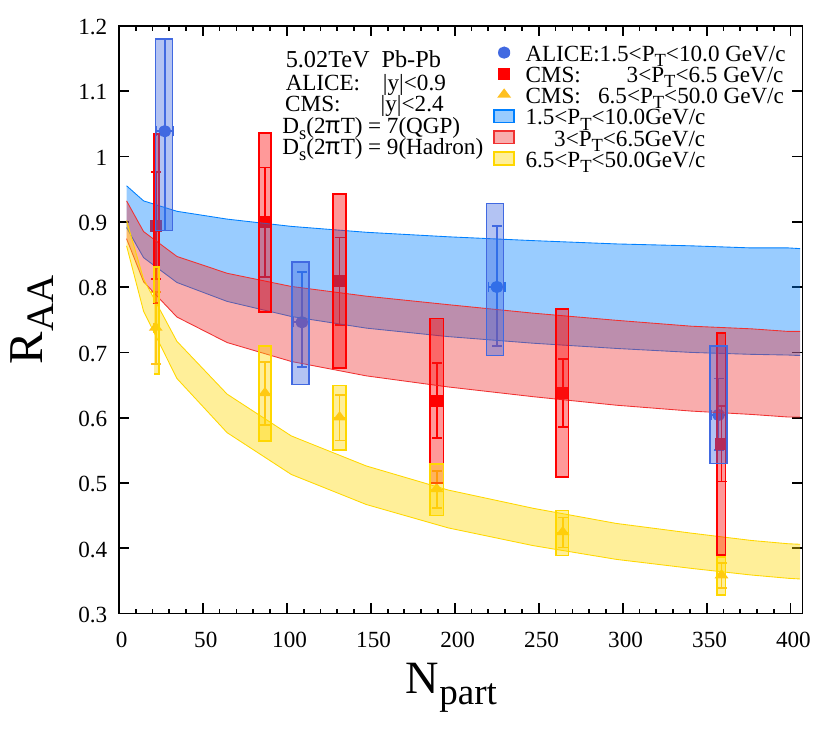}
\caption{ (Color online) Nuclear modification factor 
of non-prompt $J/\psi$ as a function of $N_{part}$ in the 
central rapidity at $\sqrt{s_{NN}}=5.02$ TeV Pb-Pb 
collisions. Different $p_T$ bins are calculated: 
$1.5<p_T<10$ GeV/c, $3<p_T<6.5$ GeV/c and $6.5<p_T<50$ GeV/c. The theoretical bands are induced by the 
uncertainty in the shadowing factors.
Experimental 
data are taken from the CMS~\cite{CMS:2017uuv} and ALICE~\cite{ALICE:2022qm} Collaborations.
}
\label{fig-JpsiNp}
\end{figure}

With hydrodynamic equations for the medium expansion and the Langevin equation for the evolution of bottom quarks and $B$ mesons, one can get the final spectrum and the nuclear modification factor $R_{AA}$ of non-prompt $J/\psi$ in Pb-Pb collisions at $\sqrt{s_{NN}}=5.02$ TeV. The spatial diffusion 
coefficient of bottom quarks and $B$ meson are taken as 
$\mathcal{D}_s(2\pi T)=7$ and $9$ respectively as explained before. 
Non-prompt $J/\psi$ $R_{AA}$ as a function of the number of participants $N_{part}$ are plotted in Fig.~\ref{fig-JpsiNp}. Bottom quarks with large 
momentum lose energy in the hot medium. They are moved to the low 
$p_T$ regions after they travel through the hot medium. This results 
in a significant suppression in the $R_{AA}$ at $6.5<p_T<50$ GeV/c. In this high $p_T$ bin, the medium-induced radiation dominates the energy loss of bottom quarks. At low $p_T$, the $R_{AA}$ becomes larger where the effect of the energy loss is not evident compared 
with the situation at high $p_T$. The shadowing effect 
reduces the bottom quark production and $R_{AA}$ at different $p_T$ bins in Fig.~\ref{fig-JpsiNp}. The band in the figure is due to the uncertainty in 
the shadowing effect. Besides, the band becomes smaller at high 
$p_T$ because both the central value and the uncertainty of the 
shadowing effect depend on $p_T$. This feature is reflected in both 
$R_{AA}(N_{part})$ and also $R_{AA}(p_T)$. As one can see 
in Fig.~\ref{fig-JpsiNp}, theoretical 
calculations qualitatively explain well the data from the CMS and 
ALICE Collaborations. 

\begin{figure}[!hbt]
\centering
\includegraphics[width=0.38\textwidth]{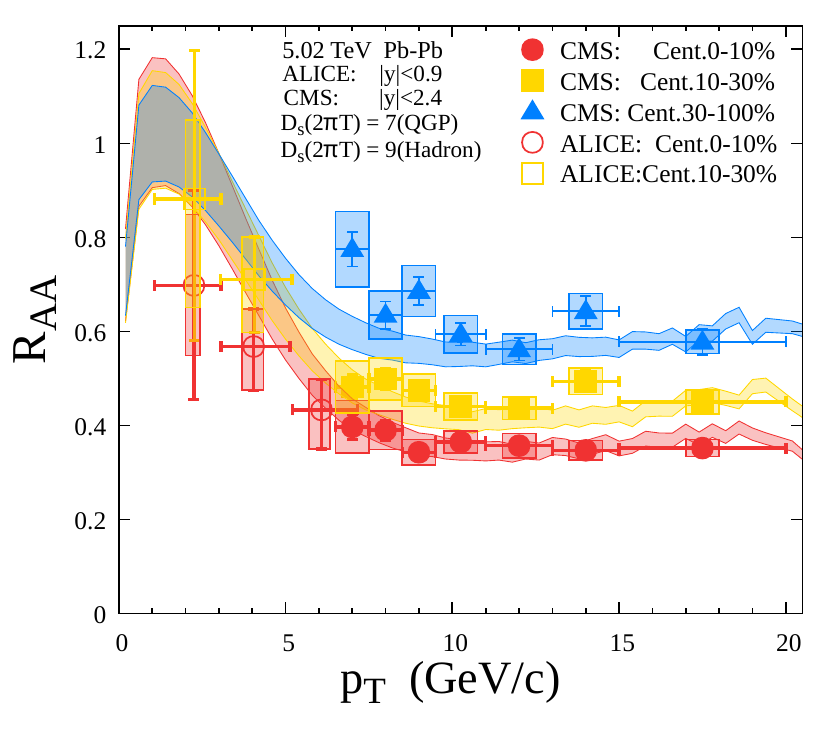}
\caption{(Color online) The nuclear modification factor of 
non-prompt $J/\psi$ as a function of $p_T$ in Pb-Pb collisions at $\sqrt{s_{NN}}=5.02$ TeV. The collision centralities are 0-10\%, 10-30\% 
and 30-100\%. The band in theoretical calculations comes 
from the uncertainty in the shadowing factor. 
Experimental 
data are taken from the CMS~\cite{CMS:2017uuv} and ALICE~\cite{ALICE:2022qm} Collaborations.
}
\label{fig-Jpsipt}
\end{figure}

\begin{figure}[!hbt]
\centering
\includegraphics[width=0.38\textwidth]{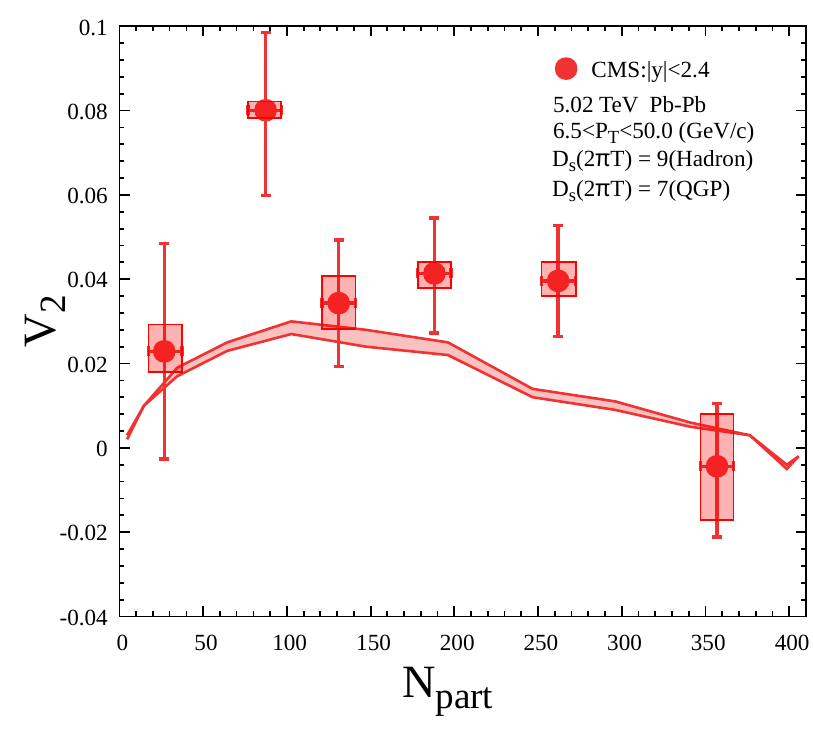}
\caption{(Color online) Elliptic flows of non-prompt 
$J/\psi$ as a function of number of participants $N_{part}$ in $\sqrt{s_{NN}}=5.02$ TeV Pb-Pb collisions. The selected $p_T$-bin is $6.5-50$ GeV/c. The spatial 
diffusion coefficient is taken to be $D_s(2\pi T)=7$ and $9$ respectively in the QGP and the hadronic gas. The theoretical band represents the uncertainty in the shadowing factor. Experimental data is cited from the CMS Collaboration~\cite{CMS:2022gvy}.}
\label{fig-v2-Np}
\end{figure}
\begin{figure}[!hbt]
\centering
\includegraphics[width=0.38\textwidth]{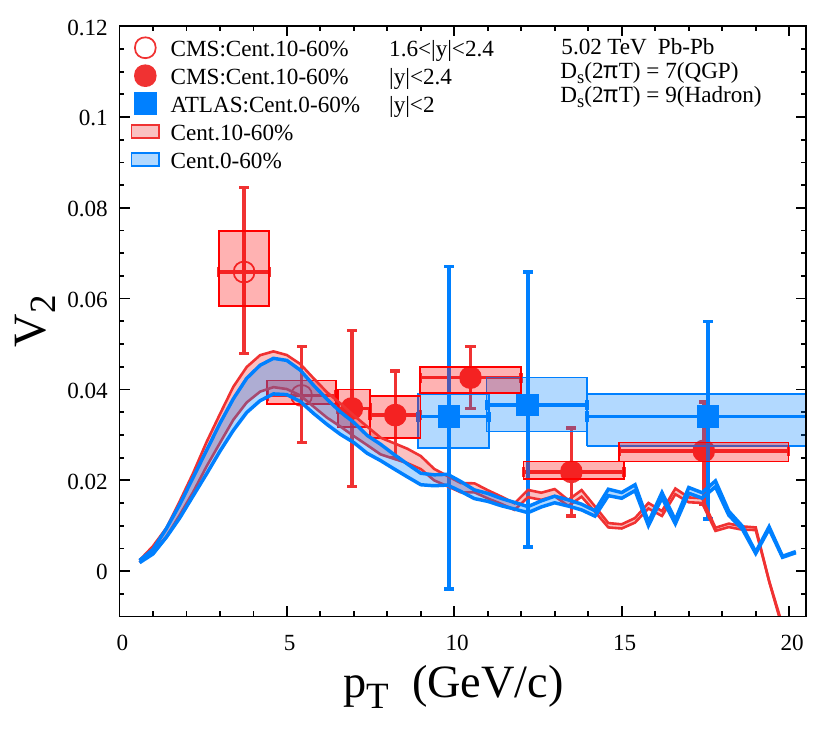}
\caption{(Color online) Elliptic flows of non-prompt 
$J/\psi$ as a function of $p_T$ in Pb-Pb collisions at $\sqrt{s_{NN}}=5.02$ TeV. The collision centralities are 
0-60\% and 10-60\%. Theoretical bands represent the uncertainty in the shadowing factor. Experimental data is cited from the CMS and ATLAS Collaborations~\cite{CMS:2022gvy,ATLAS:2018xms}. 
}
\label{fig-v2-pt}
\end{figure}

In Fig.~\ref{fig-Jpsipt}, the $p_T$-differential nuclear modification 
factor at fixed collision centralities is plotted. Bottom quarks lose energy in the hot medium. They are shifted from 
high to low $p_T$, which results in a significant suppression 
in $R_{AA}$ at high $p_T$ and a corresponding 
enhancement at low $p_T$. This modification becomes more 
evident in the central collisions compared with the peripheral collisions. 
The band in theoretical calculations is induced by the uncertainty in the shadowing 
factor. When the shadowing effect is strong, $R_{AA}$ becomes 
smaller than the unit even at low $p_T$. 

We also calculated the elliptic flows $v_2$ of non-prompt 
$J/\psi$ as a function of $N_{part}$ and $p_T$ in Pb-Pb collisions. In Fig.~\ref{fig-v2-Np}, non-prompt $J/\psi$ with high $p_T$ are selected. In the most central 
collisions, the initial energy density of the 
hot medium produced in nuclear 
collisions are nearly isotropic on 
the transverse plane when neglecting the event-by-event fluctuations. Collective expansion of the medium 
is isotropic on the transverse plane due to the same 
pressure gradients along $x$- and $y$-directions. 
In semi-central collisions, the accelerations of the 
medium become different along $x$- and $y$-directions on the transverse plane. This anisotropic expansion will 
be inherited by heavy quarks due to the strong coupling strength. 
The theoretical results are slightly below the experimental data at high $p_T$, possibly due to the absence of event-by-event fluctuations in theoretical calculations.   
In Fig.~\ref{fig-v2-pt}, elliptic flows of non-prompt $J/\psi$ as a function of transverse momentum are also calculated in centralities 0-60\% and 10-60\%. Theoretical calculations 
qualitatively explain most of the data points at middle and low $p_T$, while underestimating the data at high $p_T$ but still within 
the large error bars. In the theoretical calculations, the momentum dependence has been neglected in the value of $\mathcal{D}_s(2\pi T)$, which may underestimate the $v_2$ of heavy quarks at high $p_T$. When the value of $\mathcal{D}_s(2\pi T)$ becomes smaller at high $p_T$, the coupling strength between heavy quarks and the medium is also stronger, where $v_2$ is enhanced at high $p_T$. The detailed momentum dependence in $\mathcal{D}_s(2\pi T)$ will be left in future works.

\section{summary}
\label{sec.summary}

In this paper, we employ 
the Langevin equation to study the energy loss of bottom quarks in the quark-gluon plasma and $B$ mesons in the hadronic medium. Both contributions of elastic collisions 
and the medium-induced radiation are included in the bottom quark evolution. At 
the hadronization hypersurface, $B$ mesons are produced via the coalescence between the bottom quark and a thermal light quark, which finally decay into non-prompt $J/\psi$ after $B$ mesons move out of the 
hadronic medium. The cold nuclear matter effect is also 
considered by modifying the initial distribution of bottom quarks in the Pb-Pb collisions. The space and time evolution 
of the hot medium is given by the hydrodynamic equations. With 
realistic treatment of heavy quarks and the thermal medium, 
we calculated the nuclear modification factor and elliptic flow of non-prompt 
$J/\psi$ in different collision centralities and transverse 
momentum. Theoretical results explain the experimental 
data well with the proper values of the spatial diffusion coefficients 
determined in previous works. The spectra and elliptic flow of non-prompt $J/\psi$ reflect the energy loss of bottom quarks in the hot medium. This offers the opportunity to study the bottom quark energy loss mechanism in experiments.

\vspace{1cm}
{\bf Acknowledgement:}
Baoyi Chen appreciates inspiring discussions with Prof. Xiaozhi Bai and Prof. Pengfei Zhuang. 
This work is supported by the National Natural Science Foundation of China
(NSFC) under Grant Nos 12175165. J. Zhao is supported by funding from the European Union’s Horizon 2020 research and innovation program under grant agreement No 824093 (STRONG-2020).

\bibliography{ref}

\begin{thebibliography}{77}
\expandafter\ifx\csname natexlab\endcsname\relax\def\natexlab#1{#1}\fi
\expandafter\ifx\csname bibnamefont\endcsname\relax
  \def\bibnamefont#1{#1}\fi
\expandafter\ifx\csname bibfnamefont\endcsname\relax
  \def\bibfnamefont#1{#1}\fi
\expandafter\ifx\csname citenamefont\endcsname\relax
  \def\citenamefont#1{#1}\fi
\expandafter\ifx\csname url\endcsname\relax
  \def\url#1{\texttt{#1}}\fi
\expandafter\ifx\csname urlprefix\endcsname\relax\def\urlprefix{URL }\fi
\providecommand{\bibinfo}[2]{#2}
\providecommand{\eprint}[2][]{\url{#2}}

\bibitem[{\citenamefont{Bazavov et~al.}(2012)}]{Bazavov:2011nk}
\bibinfo{author}{\bibfnamefont{A.}~\bibnamefont{Bazavov}} \bibnamefont{et~al.},
  \bibinfo{journal}{Phys. Rev. D} \textbf{\bibinfo{volume}{85}},
  \bibinfo{pages}{054503} (\bibinfo{year}{2012}), \eprint{1111.1710}.

\bibitem[{\citenamefont{Satz}(2000)}]{Satz:2000bn}
\bibinfo{author}{\bibfnamefont{H.}~\bibnamefont{Satz}}, \bibinfo{journal}{Rept.
  Prog. Phys.} \textbf{\bibinfo{volume}{63}}, \bibinfo{pages}{1511}
  (\bibinfo{year}{2000}), \eprint{hep-ph/0007069}.

\bibitem[{\citenamefont{Bass et~al.}(1999)\citenamefont{Bass, Gyulassy,
  Stoecker, and Greiner}}]{Bass:1998vz}
\bibinfo{author}{\bibfnamefont{S.~A.} \bibnamefont{Bass}},
  \bibinfo{author}{\bibfnamefont{M.}~\bibnamefont{Gyulassy}},
  \bibinfo{author}{\bibfnamefont{H.}~\bibnamefont{Stoecker}}, \bibnamefont{and}
  \bibinfo{author}{\bibfnamefont{W.}~\bibnamefont{Greiner}},
  \bibinfo{journal}{J. Phys. G} \textbf{\bibinfo{volume}{25}},
  \bibinfo{pages}{R1} (\bibinfo{year}{1999}), \eprint{hep-ph/9810281}.

\bibitem[{\citenamefont{Cassing and Bratkovskaya}(1999)}]{Cassing:1999es}
\bibinfo{author}{\bibfnamefont{W.}~\bibnamefont{Cassing}} \bibnamefont{and}
  \bibinfo{author}{\bibfnamefont{E.~L.} \bibnamefont{Bratkovskaya}},
  \bibinfo{journal}{Phys. Rept.} \textbf{\bibinfo{volume}{308}},
  \bibinfo{pages}{65} (\bibinfo{year}{1999}).

\bibitem[{\citenamefont{Shuryak}(1984)}]{Shuryak:1983ni}
\bibinfo{author}{\bibfnamefont{E.~V.} \bibnamefont{Shuryak}},
  \bibinfo{journal}{Phys. Rept.} \textbf{\bibinfo{volume}{115}},
  \bibinfo{pages}{151} (\bibinfo{year}{1984}).

\bibitem[{\citenamefont{Matsui and Satz}(1986)}]{Matsui:1986dk}
\bibinfo{author}{\bibfnamefont{T.}~\bibnamefont{Matsui}} \bibnamefont{and}
  \bibinfo{author}{\bibfnamefont{H.}~\bibnamefont{Satz}},
  \bibinfo{journal}{Phys. Lett. B} \textbf{\bibinfo{volume}{178}},
  \bibinfo{pages}{416} (\bibinfo{year}{1986}).

\bibitem[{\citenamefont{Zhao and Rapp}(2010)}]{Zhao:2010nk}
\bibinfo{author}{\bibfnamefont{X.}~\bibnamefont{Zhao}} \bibnamefont{and}
  \bibinfo{author}{\bibfnamefont{R.}~\bibnamefont{Rapp}},
  \bibinfo{journal}{Phys. Rev. C} \textbf{\bibinfo{volume}{82}},
  \bibinfo{pages}{064905} (\bibinfo{year}{2010}), \eprint{1008.5328}.

\bibitem[{\citenamefont{Du and Rapp}(2015)}]{Du:2015wha}
\bibinfo{author}{\bibfnamefont{X.}~\bibnamefont{Du}} \bibnamefont{and}
  \bibinfo{author}{\bibfnamefont{R.}~\bibnamefont{Rapp}},
  \bibinfo{journal}{Nucl. Phys. A} \textbf{\bibinfo{volume}{943}},
  \bibinfo{pages}{147} (\bibinfo{year}{2015}), \eprint{1504.00670}.

\bibitem[{\citenamefont{Liu et~al.}(2011)\citenamefont{Liu, Chen, Xu, and
  Zhuang}}]{Liu:2010ej}
\bibinfo{author}{\bibfnamefont{Y.}~\bibnamefont{Liu}},
  \bibinfo{author}{\bibfnamefont{B.}~\bibnamefont{Chen}},
  \bibinfo{author}{\bibfnamefont{N.}~\bibnamefont{Xu}}, \bibnamefont{and}
  \bibinfo{author}{\bibfnamefont{P.}~\bibnamefont{Zhuang}},
  \bibinfo{journal}{Phys. Lett. B} \textbf{\bibinfo{volume}{697}},
  \bibinfo{pages}{32} (\bibinfo{year}{2011}), \eprint{1009.2585}.

\bibitem[{\citenamefont{Zhou et~al.}(2014)\citenamefont{Zhou, Xu, Xu, and
  Zhuang}}]{Zhou:2014kka}
\bibinfo{author}{\bibfnamefont{K.}~\bibnamefont{Zhou}},
  \bibinfo{author}{\bibfnamefont{N.}~\bibnamefont{Xu}},
  \bibinfo{author}{\bibfnamefont{Z.}~\bibnamefont{Xu}}, \bibnamefont{and}
  \bibinfo{author}{\bibfnamefont{P.}~\bibnamefont{Zhuang}},
  \bibinfo{journal}{Phys. Rev. C} \textbf{\bibinfo{volume}{89}},
  \bibinfo{pages}{054911} (\bibinfo{year}{2014}), \eprint{1401.5845}.

\bibitem[{\citenamefont{Chen}(2019)}]{Chen:2018kfo}
\bibinfo{author}{\bibfnamefont{B.}~\bibnamefont{Chen}}, \bibinfo{journal}{Chin.
  Phys. C} \textbf{\bibinfo{volume}{43}}, \bibinfo{pages}{124101}
  (\bibinfo{year}{2019}), \eprint{1811.11393}.

\bibitem[{\citenamefont{Zhao and Chen}(2018)}]{Zhao:2017yan}
\bibinfo{author}{\bibfnamefont{J.}~\bibnamefont{Zhao}} \bibnamefont{and}
  \bibinfo{author}{\bibfnamefont{B.}~\bibnamefont{Chen}},
  \bibinfo{journal}{Phys. Lett. B} \textbf{\bibinfo{volume}{776}},
  \bibinfo{pages}{17} (\bibinfo{year}{2018}), \eprint{1705.04558}.

\bibitem[{\citenamefont{Yao and Mehen}(2021)}]{Yao:2020eqy}
\bibinfo{author}{\bibfnamefont{X.}~\bibnamefont{Yao}} \bibnamefont{and}
  \bibinfo{author}{\bibfnamefont{T.}~\bibnamefont{Mehen}},
  \bibinfo{journal}{JHEP} \textbf{\bibinfo{volume}{02}}, \bibinfo{pages}{062}
  (\bibinfo{year}{2021}), \eprint{2009.02408}.

\bibitem[{\citenamefont{Blaizot et~al.}(2016)\citenamefont{Blaizot, De~Boni,
  Faccioli, and Garberoglio}}]{Blaizot:2015hya}
\bibinfo{author}{\bibfnamefont{J.-P.} \bibnamefont{Blaizot}},
  \bibinfo{author}{\bibfnamefont{D.}~\bibnamefont{De~Boni}},
  \bibinfo{author}{\bibfnamefont{P.}~\bibnamefont{Faccioli}}, \bibnamefont{and}
  \bibinfo{author}{\bibfnamefont{G.}~\bibnamefont{Garberoglio}},
  \bibinfo{journal}{Nucl. Phys. A} \textbf{\bibinfo{volume}{946}},
  \bibinfo{pages}{49} (\bibinfo{year}{2016}), \eprint{1503.03857}.

\bibitem[{\citenamefont{Brambilla et~al.}(2021)\citenamefont{Brambilla,
  Escobedo, Strickland, Vairo, Vander~Griend, and Weber}}]{Brambilla:2020qwo}
\bibinfo{author}{\bibfnamefont{N.}~\bibnamefont{Brambilla}},
  \bibinfo{author}{\bibfnamefont{M.~A.} \bibnamefont{Escobedo}},
  \bibinfo{author}{\bibfnamefont{M.}~\bibnamefont{Strickland}},
  \bibinfo{author}{\bibfnamefont{A.}~\bibnamefont{Vairo}},
  \bibinfo{author}{\bibfnamefont{P.}~\bibnamefont{Vander~Griend}},
  \bibnamefont{and} \bibinfo{author}{\bibfnamefont{J.~H.} \bibnamefont{Weber}},
  \bibinfo{journal}{JHEP} \textbf{\bibinfo{volume}{05}}, \bibinfo{pages}{136}
  (\bibinfo{year}{2021}), \eprint{2012.01240}.

\bibitem[{\citenamefont{Delorme et~al.}(2022)\citenamefont{Delorme, Gousset,
  Katz, and Gossiaux}}]{Delorme:2022hoo}
\bibinfo{author}{\bibfnamefont{S.}~\bibnamefont{Delorme}},
  \bibinfo{author}{\bibfnamefont{T.}~\bibnamefont{Gousset}},
  \bibinfo{author}{\bibfnamefont{R.}~\bibnamefont{Katz}}, \bibnamefont{and}
  \bibinfo{author}{\bibfnamefont{P.-B.} \bibnamefont{Gossiaux}},
  \bibinfo{journal}{EPJ Web Conf.} \textbf{\bibinfo{volume}{258}},
  \bibinfo{pages}{05009} (\bibinfo{year}{2022}).

\bibitem[{\citenamefont{Miura et~al.}(2022)\citenamefont{Miura, Akamatsu,
  Asakawa, and Kaida}}]{Miura:2022arv}
\bibinfo{author}{\bibfnamefont{T.}~\bibnamefont{Miura}},
  \bibinfo{author}{\bibfnamefont{Y.}~\bibnamefont{Akamatsu}},
  \bibinfo{author}{\bibfnamefont{M.}~\bibnamefont{Asakawa}}, \bibnamefont{and}
  \bibinfo{author}{\bibfnamefont{Y.}~\bibnamefont{Kaida}},
  \bibinfo{journal}{Phys. Rev. D} \textbf{\bibinfo{volume}{106}},
  \bibinfo{pages}{074001} (\bibinfo{year}{2022}), \eprint{2205.15551}.

\bibitem[{\citenamefont{Zha et~al.}(2019)\citenamefont{Zha, Ruan, Tang, Xu, and
  Yang}}]{Zha:2018ytv}
\bibinfo{author}{\bibfnamefont{W.}~\bibnamefont{Zha}},
  \bibinfo{author}{\bibfnamefont{L.}~\bibnamefont{Ruan}},
  \bibinfo{author}{\bibfnamefont{Z.}~\bibnamefont{Tang}},
  \bibinfo{author}{\bibfnamefont{Z.}~\bibnamefont{Xu}}, \bibnamefont{and}
  \bibinfo{author}{\bibfnamefont{S.}~\bibnamefont{Yang}},
  \bibinfo{journal}{Phys. Lett. B} \textbf{\bibinfo{volume}{789}},
  \bibinfo{pages}{238} (\bibinfo{year}{2019}), \eprint{1810.02064}.

\bibitem[{\citenamefont{Shi et~al.}(2018)\citenamefont{Shi, Zha, and
  Chen}}]{Shi:2017qep}
\bibinfo{author}{\bibfnamefont{W.}~\bibnamefont{Shi}},
  \bibinfo{author}{\bibfnamefont{W.}~\bibnamefont{Zha}}, \bibnamefont{and}
  \bibinfo{author}{\bibfnamefont{B.}~\bibnamefont{Chen}},
  \bibinfo{journal}{Phys. Lett. B} \textbf{\bibinfo{volume}{777}},
  \bibinfo{pages}{399} (\bibinfo{year}{2018}), \eprint{1710.00332}.

\bibitem[{\citenamefont{Zhao et~al.}(2022)\citenamefont{Zhao, Chen, and
  Zhuang}}]{Zhao:2021voa}
\bibinfo{author}{\bibfnamefont{J.}~\bibnamefont{Zhao}},
  \bibinfo{author}{\bibfnamefont{B.}~\bibnamefont{Chen}}, \bibnamefont{and}
  \bibinfo{author}{\bibfnamefont{P.}~\bibnamefont{Zhuang}},
  \bibinfo{journal}{Phys. Rev. C} \textbf{\bibinfo{volume}{105}},
  \bibinfo{pages}{034902} (\bibinfo{year}{2022}), \eprint{2112.00293}.

\bibitem[{\citenamefont{Chen et~al.}(2020)\citenamefont{Chen, Hu, Zhang, and
  Zhao}}]{Chen:2019qzx}
\bibinfo{author}{\bibfnamefont{B.}~\bibnamefont{Chen}},
  \bibinfo{author}{\bibfnamefont{M.}~\bibnamefont{Hu}},
  \bibinfo{author}{\bibfnamefont{H.}~\bibnamefont{Zhang}}, \bibnamefont{and}
  \bibinfo{author}{\bibfnamefont{J.}~\bibnamefont{Zhao}},
  \bibinfo{journal}{Phys. Lett. B} \textbf{\bibinfo{volume}{802}},
  \bibinfo{pages}{135271} (\bibinfo{year}{2020}), \eprint{1910.08275}.

\bibitem[{\citenamefont{Rothkopf}(2020)}]{Rothkopf:2019ipj}
\bibinfo{author}{\bibfnamefont{A.}~\bibnamefont{Rothkopf}},
  \bibinfo{journal}{Phys. Rept.} \textbf{\bibinfo{volume}{858}},
  \bibinfo{pages}{1} (\bibinfo{year}{2020}), \eprint{1912.02253}.

\bibitem[{\citenamefont{Zhao et~al.}(2020)\citenamefont{Zhao, Zhou, Chen, and
  Zhuang}}]{Zhao:2020jqu}
\bibinfo{author}{\bibfnamefont{J.}~\bibnamefont{Zhao}},
  \bibinfo{author}{\bibfnamefont{K.}~\bibnamefont{Zhou}},
  \bibinfo{author}{\bibfnamefont{S.}~\bibnamefont{Chen}}, \bibnamefont{and}
  \bibinfo{author}{\bibfnamefont{P.}~\bibnamefont{Zhuang}},
  \bibinfo{journal}{Prog. Part. Nucl. Phys.} \textbf{\bibinfo{volume}{114}},
  \bibinfo{pages}{103801} (\bibinfo{year}{2020}), \eprint{2005.08277}.

\bibitem[{\citenamefont{Djordjevic and Gyulassy}(2004)}]{Djordjevic:2003zk}
\bibinfo{author}{\bibfnamefont{M.}~\bibnamefont{Djordjevic}} \bibnamefont{and}
  \bibinfo{author}{\bibfnamefont{M.}~\bibnamefont{Gyulassy}},
  \bibinfo{journal}{Nucl. Phys. A} \textbf{\bibinfo{volume}{733}},
  \bibinfo{pages}{265} (\bibinfo{year}{2004}), \eprint{nucl-th/0310076}.

\bibitem[{\citenamefont{Wicks et~al.}(2007)\citenamefont{Wicks, Horowitz,
  Djordjevic, and Gyulassy}}]{Wicks:2005gt}
\bibinfo{author}{\bibfnamefont{S.}~\bibnamefont{Wicks}},
  \bibinfo{author}{\bibfnamefont{W.}~\bibnamefont{Horowitz}},
  \bibinfo{author}{\bibfnamefont{M.}~\bibnamefont{Djordjevic}},
  \bibnamefont{and} \bibinfo{author}{\bibfnamefont{M.}~\bibnamefont{Gyulassy}},
  \bibinfo{journal}{Nucl. Phys. A} \textbf{\bibinfo{volume}{784}},
  \bibinfo{pages}{426} (\bibinfo{year}{2007}), \eprint{nucl-th/0512076}.

\bibitem[{\citenamefont{Braaten and Thoma}(1991)}]{Braaten:1991we}
\bibinfo{author}{\bibfnamefont{E.}~\bibnamefont{Braaten}} \bibnamefont{and}
  \bibinfo{author}{\bibfnamefont{M.~H.} \bibnamefont{Thoma}},
  \bibinfo{journal}{Phys. Rev. D} \textbf{\bibinfo{volume}{44}},
  \bibinfo{pages}{R2625} (\bibinfo{year}{1991}).

\bibitem[{\citenamefont{Moore and Teaney}(2005)}]{Moore:2004tg}
\bibinfo{author}{\bibfnamefont{G.~D.} \bibnamefont{Moore}} \bibnamefont{and}
  \bibinfo{author}{\bibfnamefont{D.}~\bibnamefont{Teaney}},
  \bibinfo{journal}{Phys. Rev. C} \textbf{\bibinfo{volume}{71}},
  \bibinfo{pages}{064904} (\bibinfo{year}{2005}), \eprint{hep-ph/0412346}.

\bibitem[{\citenamefont{Gyulassy et~al.}(2000)\citenamefont{Gyulassy, Levai,
  and Vitev}}]{Gyulassy:2000fs}
\bibinfo{author}{\bibfnamefont{M.}~\bibnamefont{Gyulassy}},
  \bibinfo{author}{\bibfnamefont{P.}~\bibnamefont{Levai}}, \bibnamefont{and}
  \bibinfo{author}{\bibfnamefont{I.}~\bibnamefont{Vitev}},
  \bibinfo{journal}{Phys. Rev. Lett.} \textbf{\bibinfo{volume}{85}},
  \bibinfo{pages}{5535} (\bibinfo{year}{2000}), \eprint{nucl-th/0005032}.

\bibitem[{\citenamefont{Arnold et~al.}(2003)\citenamefont{Arnold, Moore, and
  Yaffe}}]{Arnold:2002zm}
\bibinfo{author}{\bibfnamefont{P.~B.} \bibnamefont{Arnold}},
  \bibinfo{author}{\bibfnamefont{G.~D.} \bibnamefont{Moore}}, \bibnamefont{and}
  \bibinfo{author}{\bibfnamefont{L.~G.} \bibnamefont{Yaffe}},
  \bibinfo{journal}{JHEP} \textbf{\bibinfo{volume}{01}}, \bibinfo{pages}{030}
  (\bibinfo{year}{2003}), \eprint{hep-ph/0209353}.

\bibitem[{\citenamefont{Zhang et~al.}(2004)\citenamefont{Zhang, Wang, and
  Wang}}]{Zhang:2003wk}
\bibinfo{author}{\bibfnamefont{B.-W.} \bibnamefont{Zhang}},
  \bibinfo{author}{\bibfnamefont{E.}~\bibnamefont{Wang}}, \bibnamefont{and}
  \bibinfo{author}{\bibfnamefont{X.-N.} \bibnamefont{Wang}},
  \bibinfo{journal}{Phys. Rev. Lett.} \textbf{\bibinfo{volume}{93}},
  \bibinfo{pages}{072301} (\bibinfo{year}{2004}), \eprint{nucl-th/0309040}.

\bibitem[{\citenamefont{Qin et~al.}(2008)\citenamefont{Qin, Ruppert, Gale,
  Jeon, Moore, and Mustafa}}]{Qin:2007rn}
\bibinfo{author}{\bibfnamefont{G.-Y.} \bibnamefont{Qin}},
  \bibinfo{author}{\bibfnamefont{J.}~\bibnamefont{Ruppert}},
  \bibinfo{author}{\bibfnamefont{C.}~\bibnamefont{Gale}},
  \bibinfo{author}{\bibfnamefont{S.}~\bibnamefont{Jeon}},
  \bibinfo{author}{\bibfnamefont{G.~D.} \bibnamefont{Moore}}, \bibnamefont{and}
  \bibinfo{author}{\bibfnamefont{M.~G.} \bibnamefont{Mustafa}},
  \bibinfo{journal}{Phys. Rev. Lett.} \textbf{\bibinfo{volume}{100}},
  \bibinfo{pages}{072301} (\bibinfo{year}{2008}), \eprint{0710.0605}.

\bibitem[{\citenamefont{Majumder}(2012)}]{Majumder:2009ge}
\bibinfo{author}{\bibfnamefont{A.}~\bibnamefont{Majumder}},
  \bibinfo{journal}{Phys. Rev. D} \textbf{\bibinfo{volume}{85}},
  \bibinfo{pages}{014023} (\bibinfo{year}{2012}), \eprint{0912.2987}.

\bibitem[{\citenamefont{Berrehrah et~al.}(2014)\citenamefont{Berrehrah,
  Gossiaux, Aichelin, Cassing, and Bratkovskaya}}]{Berrehrah:2014kba}
\bibinfo{author}{\bibfnamefont{H.}~\bibnamefont{Berrehrah}},
  \bibinfo{author}{\bibfnamefont{P.-B.} \bibnamefont{Gossiaux}},
  \bibinfo{author}{\bibfnamefont{J.}~\bibnamefont{Aichelin}},
  \bibinfo{author}{\bibfnamefont{W.}~\bibnamefont{Cassing}}, \bibnamefont{and}
  \bibinfo{author}{\bibfnamefont{E.}~\bibnamefont{Bratkovskaya}},
  \bibinfo{journal}{Phys. Rev. C} \textbf{\bibinfo{volume}{90}},
  \bibinfo{pages}{064906} (\bibinfo{year}{2014}), \eprint{1405.3243}.

\bibitem[{\citenamefont{Cao et~al.}(2019)}]{Cao:2018ews}
\bibinfo{author}{\bibfnamefont{S.}~\bibnamefont{Cao}} \bibnamefont{et~al.},
  \bibinfo{journal}{Phys. Rev. C} \textbf{\bibinfo{volume}{99}},
  \bibinfo{pages}{054907} (\bibinfo{year}{2019}), \eprint{1809.07894}.

\bibitem[{\citenamefont{Beraudo et~al.}(2018)}]{Rapp:2018qla}
\bibinfo{author}{\bibfnamefont{A.}~\bibnamefont{Beraudo}} \bibnamefont{et~al.},
  \bibinfo{journal}{Nucl. Phys. A} \textbf{\bibinfo{volume}{979}},
  \bibinfo{pages}{21} (\bibinfo{year}{2018}), \eprint{1803.03824}.

\bibitem[{\citenamefont{Gossiaux et~al.}(2010)\citenamefont{Gossiaux, Aichelin,
  Gousset, and Guiho}}]{Gossiaux:2010yx}
\bibinfo{author}{\bibfnamefont{P.~B.} \bibnamefont{Gossiaux}},
  \bibinfo{author}{\bibfnamefont{J.}~\bibnamefont{Aichelin}},
  \bibinfo{author}{\bibfnamefont{T.}~\bibnamefont{Gousset}}, \bibnamefont{and}
  \bibinfo{author}{\bibfnamefont{V.}~\bibnamefont{Guiho}}, \bibinfo{journal}{J.
  Phys. G} \textbf{\bibinfo{volume}{37}}, \bibinfo{pages}{094019}
  (\bibinfo{year}{2010}), \eprint{1001.4166}.

\bibitem[{\citenamefont{He et~al.}(2013)\citenamefont{He, Fries, and
  Rapp}}]{He:2012df}
\bibinfo{author}{\bibfnamefont{M.}~\bibnamefont{He}},
  \bibinfo{author}{\bibfnamefont{R.~J.} \bibnamefont{Fries}}, \bibnamefont{and}
  \bibinfo{author}{\bibfnamefont{R.}~\bibnamefont{Rapp}},
  \bibinfo{journal}{Phys. Rev. Lett.} \textbf{\bibinfo{volume}{110}},
  \bibinfo{pages}{112301} (\bibinfo{year}{2013}), \eprint{1204.4442}.

\bibitem[{\citenamefont{Das et~al.}(2013)\citenamefont{Das, Chandra, and
  Alam}}]{Das:2012ck}
\bibinfo{author}{\bibfnamefont{S.~K.} \bibnamefont{Das}},
  \bibinfo{author}{\bibfnamefont{V.}~\bibnamefont{Chandra}}, \bibnamefont{and}
  \bibinfo{author}{\bibfnamefont{J.-e.} \bibnamefont{Alam}},
  \bibinfo{journal}{J. Phys. G} \textbf{\bibinfo{volume}{41}},
  \bibinfo{pages}{015102} (\bibinfo{year}{2013}), \eprint{1210.3905}.

\bibitem[{\citenamefont{Mazumder et~al.}(2014)\citenamefont{Mazumder,
  Bhattacharyya, and Alam}}]{Mazumder:2013oaa}
\bibinfo{author}{\bibfnamefont{S.}~\bibnamefont{Mazumder}},
  \bibinfo{author}{\bibfnamefont{T.}~\bibnamefont{Bhattacharyya}},
  \bibnamefont{and} \bibinfo{author}{\bibfnamefont{J.-e.} \bibnamefont{Alam}},
  \bibinfo{journal}{Phys. Rev. D} \textbf{\bibinfo{volume}{89}},
  \bibinfo{pages}{014002} (\bibinfo{year}{2014}), \eprint{1305.6445}.

\bibitem[{\citenamefont{Song et~al.}(2020)\citenamefont{Song, Moreau, Aichelin,
  and Bratkovskaya}}]{Song:2019cqz}
\bibinfo{author}{\bibfnamefont{T.}~\bibnamefont{Song}},
  \bibinfo{author}{\bibfnamefont{P.}~\bibnamefont{Moreau}},
  \bibinfo{author}{\bibfnamefont{J.}~\bibnamefont{Aichelin}}, \bibnamefont{and}
  \bibinfo{author}{\bibfnamefont{E.}~\bibnamefont{Bratkovskaya}},
  \bibinfo{journal}{Phys. Rev. C} \textbf{\bibinfo{volume}{101}},
  \bibinfo{pages}{044901} (\bibinfo{year}{2020}), \eprint{1910.09889}.

\bibitem[{\citenamefont{Kurian et~al.}(2020)\citenamefont{Kurian, Singh,
  Chandra, Jeon, and Gale}}]{Kurian:2020orp}
\bibinfo{author}{\bibfnamefont{M.}~\bibnamefont{Kurian}},
  \bibinfo{author}{\bibfnamefont{M.}~\bibnamefont{Singh}},
  \bibinfo{author}{\bibfnamefont{V.}~\bibnamefont{Chandra}},
  \bibinfo{author}{\bibfnamefont{S.}~\bibnamefont{Jeon}}, \bibnamefont{and}
  \bibinfo{author}{\bibfnamefont{C.}~\bibnamefont{Gale}},
  \bibinfo{journal}{Phys. Rev. C} \textbf{\bibinfo{volume}{102}},
  \bibinfo{pages}{044907} (\bibinfo{year}{2020}), \eprint{2007.07705}.

\bibitem[{\citenamefont{Prakash et~al.}(2021)\citenamefont{Prakash, Kurian,
  Das, and Chandra}}]{Prakash:2021lwt}
\bibinfo{author}{\bibfnamefont{J.}~\bibnamefont{Prakash}},
  \bibinfo{author}{\bibfnamefont{M.}~\bibnamefont{Kurian}},
  \bibinfo{author}{\bibfnamefont{S.~K.} \bibnamefont{Das}}, \bibnamefont{and}
  \bibinfo{author}{\bibfnamefont{V.}~\bibnamefont{Chandra}},
  \bibinfo{journal}{Phys. Rev. D} \textbf{\bibinfo{volume}{103}},
  \bibinfo{pages}{094009} (\bibinfo{year}{2021}), \eprint{2102.07082}.

\bibitem[{\citenamefont{Cao et~al.}(2013)\citenamefont{Cao, Qin, and
  Bass}}]{Cao:2013ita}
\bibinfo{author}{\bibfnamefont{S.}~\bibnamefont{Cao}},
  \bibinfo{author}{\bibfnamefont{G.-Y.} \bibnamefont{Qin}}, \bibnamefont{and}
  \bibinfo{author}{\bibfnamefont{S.~A.} \bibnamefont{Bass}},
  \bibinfo{journal}{Phys. Rev. C} \textbf{\bibinfo{volume}{88}},
  \bibinfo{pages}{044907} (\bibinfo{year}{2013}), \eprint{1308.0617}.

\bibitem[{\citenamefont{He and Rapp}(2020)}]{He:2019vgs}
\bibinfo{author}{\bibfnamefont{M.}~\bibnamefont{He}} \bibnamefont{and}
  \bibinfo{author}{\bibfnamefont{R.}~\bibnamefont{Rapp}},
  \bibinfo{journal}{Phys. Rev. Lett.} \textbf{\bibinfo{volume}{124}},
  \bibinfo{pages}{042301} (\bibinfo{year}{2020}), \eprint{1905.09216}.

\bibitem[{\citenamefont{Gossiaux et~al.}(2009)\citenamefont{Gossiaux,
  Bierkandt, and Aichelin}}]{Gossiaux:2009mk}
\bibinfo{author}{\bibfnamefont{P.~B.} \bibnamefont{Gossiaux}},
  \bibinfo{author}{\bibfnamefont{R.}~\bibnamefont{Bierkandt}},
  \bibnamefont{and} \bibinfo{author}{\bibfnamefont{J.}~\bibnamefont{Aichelin}},
  \bibinfo{journal}{Phys. Rev. C} \textbf{\bibinfo{volume}{79}},
  \bibinfo{pages}{044906} (\bibinfo{year}{2009}), \eprint{0901.0946}.

\bibitem[{\citenamefont{Plumari et~al.}(2018)\citenamefont{Plumari, Minissale,
  Das, Coci, and Greco}}]{Plumari:2017ntm}
\bibinfo{author}{\bibfnamefont{S.}~\bibnamefont{Plumari}},
  \bibinfo{author}{\bibfnamefont{V.}~\bibnamefont{Minissale}},
  \bibinfo{author}{\bibfnamefont{S.~K.} \bibnamefont{Das}},
  \bibinfo{author}{\bibfnamefont{G.}~\bibnamefont{Coci}}, \bibnamefont{and}
  \bibinfo{author}{\bibfnamefont{V.}~\bibnamefont{Greco}},
  \bibinfo{journal}{Eur. Phys. J. C} \textbf{\bibinfo{volume}{78}},
  \bibinfo{pages}{348} (\bibinfo{year}{2018}), \eprint{1712.00730}.

\bibitem[{\citenamefont{Fuchs et~al.}(2006)\citenamefont{Fuchs, Martemyanov,
  Faessler, and Krivoruchenko}}]{Fuchs:2004fh}
\bibinfo{author}{\bibfnamefont{C.}~\bibnamefont{Fuchs}},
  \bibinfo{author}{\bibfnamefont{B.~V.} \bibnamefont{Martemyanov}},
  \bibinfo{author}{\bibfnamefont{A.}~\bibnamefont{Faessler}}, \bibnamefont{and}
  \bibinfo{author}{\bibfnamefont{M.~I.} \bibnamefont{Krivoruchenko}},
  \bibinfo{journal}{Phys. Rev. C} \textbf{\bibinfo{volume}{73}},
  \bibinfo{pages}{035204} (\bibinfo{year}{2006}), \eprint{nucl-th/0410065}.

\bibitem[{\citenamefont{He et~al.}(2011)\citenamefont{He, Fries, and
  Rapp}}]{He:2011yi}
\bibinfo{author}{\bibfnamefont{M.}~\bibnamefont{He}},
  \bibinfo{author}{\bibfnamefont{R.~J.} \bibnamefont{Fries}}, \bibnamefont{and}
  \bibinfo{author}{\bibfnamefont{R.}~\bibnamefont{Rapp}},
  \bibinfo{journal}{Phys. Lett. B} \textbf{\bibinfo{volume}{701}},
  \bibinfo{pages}{445} (\bibinfo{year}{2011}), \eprint{1103.6279}.

\bibitem[{\citenamefont{Scardina et~al.}(2017)\citenamefont{Scardina, Das,
  Minissale, Plumari, and Greco}}]{Scardina:2017ipo}
\bibinfo{author}{\bibfnamefont{F.}~\bibnamefont{Scardina}},
  \bibinfo{author}{\bibfnamefont{S.~K.} \bibnamefont{Das}},
  \bibinfo{author}{\bibfnamefont{V.}~\bibnamefont{Minissale}},
  \bibinfo{author}{\bibfnamefont{S.}~\bibnamefont{Plumari}}, \bibnamefont{and}
  \bibinfo{author}{\bibfnamefont{V.}~\bibnamefont{Greco}},
  \bibinfo{journal}{Phys. Rev. C} \textbf{\bibinfo{volume}{96}},
  \bibinfo{pages}{044905} (\bibinfo{year}{2017}), \eprint{1707.05452}.

\bibitem[{\citenamefont{Dong and Greco}(2019)}]{Dong:2019unq}
\bibinfo{author}{\bibfnamefont{X.}~\bibnamefont{Dong}} \bibnamefont{and}
  \bibinfo{author}{\bibfnamefont{V.}~\bibnamefont{Greco}},
  \bibinfo{journal}{Prog. Part. Nucl. Phys.} \textbf{\bibinfo{volume}{104}},
  \bibinfo{pages}{97} (\bibinfo{year}{2019}).

\bibitem[{\citenamefont{Abelev et~al.}(2013)}]{ALICE:2013olq}
\bibinfo{author}{\bibfnamefont{B.}~\bibnamefont{Abelev}} \bibnamefont{et~al.}
  (\bibinfo{collaboration}{ALICE}), \bibinfo{journal}{Phys. Rev. Lett.}
  \textbf{\bibinfo{volume}{111}}, \bibinfo{pages}{102301}
  (\bibinfo{year}{2013}), \eprint{1305.2707}.

\bibitem[{\citenamefont{Acharya et~al.}(2021)}]{ALICE:2020iug}
\bibinfo{author}{\bibfnamefont{S.}~\bibnamefont{Acharya}} \bibnamefont{et~al.}
  (\bibinfo{collaboration}{ALICE}), \bibinfo{journal}{Phys. Lett. B}
  \textbf{\bibinfo{volume}{813}}, \bibinfo{pages}{136054}
  (\bibinfo{year}{2021}), \eprint{2005.11131}.

\bibitem[{\citenamefont{Sirunyan et~al.}(2021)}]{CMS:2020bnz}
\bibinfo{author}{\bibfnamefont{A.~M.} \bibnamefont{Sirunyan}}
  \bibnamefont{et~al.} (\bibinfo{collaboration}{CMS}), \bibinfo{journal}{Phys.
  Lett. B} \textbf{\bibinfo{volume}{816}}, \bibinfo{pages}{136253}
  (\bibinfo{year}{2021}), \eprint{2009.12628}.

\bibitem[{\citenamefont{Sirunyan et~al.}(2018)}]{CMS:2017uuv}
\bibinfo{author}{\bibfnamefont{A.~M.} \bibnamefont{Sirunyan}}
  \bibnamefont{et~al.} (\bibinfo{collaboration}{CMS}), \bibinfo{journal}{Eur.
  Phys. J. C} \textbf{\bibinfo{volume}{78}}, \bibinfo{pages}{509}
  (\bibinfo{year}{2018}), \eprint{1712.08959}.

\bibitem[{\citenamefont{Aaboud et~al.}(2018{\natexlab{a}})}]{ATLAS:2018xms}
\bibinfo{author}{\bibfnamefont{M.}~\bibnamefont{Aaboud}} \bibnamefont{et~al.}
  (\bibinfo{collaboration}{ATLAS}), \bibinfo{journal}{Eur. Phys. J. C}
  \textbf{\bibinfo{volume}{78}}, \bibinfo{pages}{784}
  (\bibinfo{year}{2018}{\natexlab{a}}), \eprint{1807.05198}.

\bibitem[{\citenamefont{Eskola et~al.}(2009)\citenamefont{Eskola, Paukkunen,
  and Salgado}}]{Eskola:2009uj}
\bibinfo{author}{\bibfnamefont{K.~J.} \bibnamefont{Eskola}},
  \bibinfo{author}{\bibfnamefont{H.}~\bibnamefont{Paukkunen}},
  \bibnamefont{and} \bibinfo{author}{\bibfnamefont{C.~A.}
  \bibnamefont{Salgado}}, \bibinfo{journal}{JHEP}
  \textbf{\bibinfo{volume}{04}}, \bibinfo{pages}{065} (\bibinfo{year}{2009}),
  \eprint{0902.4154}.

\bibitem[{\citenamefont{Cacciari et~al.}(1998)\citenamefont{Cacciari, Greco,
  and Nason}}]{Cacciari:1998it}
\bibinfo{author}{\bibfnamefont{M.}~\bibnamefont{Cacciari}},
  \bibinfo{author}{\bibfnamefont{M.}~\bibnamefont{Greco}}, \bibnamefont{and}
  \bibinfo{author}{\bibfnamefont{P.}~\bibnamefont{Nason}},
  \bibinfo{journal}{JHEP} \textbf{\bibinfo{volume}{05}}, \bibinfo{pages}{007}
  (\bibinfo{year}{1998}), \eprint{hep-ph/9803400}.

\bibitem[{\citenamefont{Cacciari et~al.}(2012)\citenamefont{Cacciari, Frixione,
  Houdeau, Mangano, Nason, and Ridolfi}}]{Cacciari:2012ny}
\bibinfo{author}{\bibfnamefont{M.}~\bibnamefont{Cacciari}},
  \bibinfo{author}{\bibfnamefont{S.}~\bibnamefont{Frixione}},
  \bibinfo{author}{\bibfnamefont{N.}~\bibnamefont{Houdeau}},
  \bibinfo{author}{\bibfnamefont{M.~L.} \bibnamefont{Mangano}},
  \bibinfo{author}{\bibfnamefont{P.}~\bibnamefont{Nason}}, \bibnamefont{and}
  \bibinfo{author}{\bibfnamefont{G.}~\bibnamefont{Ridolfi}},
  \bibinfo{journal}{JHEP} \textbf{\bibinfo{volume}{10}}, \bibinfo{pages}{137}
  (\bibinfo{year}{2012}), \eprint{1205.6344}.

\bibitem[{\citenamefont{Chen and Zhao}(2017)}]{Chen:2017duy}
\bibinfo{author}{\bibfnamefont{B.}~\bibnamefont{Chen}} \bibnamefont{and}
  \bibinfo{author}{\bibfnamefont{J.}~\bibnamefont{Zhao}},
  \bibinfo{journal}{Phys. Lett. B} \textbf{\bibinfo{volume}{772}},
  \bibinfo{pages}{819} (\bibinfo{year}{2017}), \eprint{1704.05622}.

\bibitem[{\citenamefont{Das et~al.}(2015)\citenamefont{Das, Scardina, Plumari,
  and Greco}}]{Das:2015ana}
\bibinfo{author}{\bibfnamefont{S.~K.} \bibnamefont{Das}},
  \bibinfo{author}{\bibfnamefont{F.}~\bibnamefont{Scardina}},
  \bibinfo{author}{\bibfnamefont{S.}~\bibnamefont{Plumari}}, \bibnamefont{and}
  \bibinfo{author}{\bibfnamefont{V.}~\bibnamefont{Greco}},
  \bibinfo{journal}{Phys. Lett. B} \textbf{\bibinfo{volume}{747}},
  \bibinfo{pages}{260} (\bibinfo{year}{2015}), \eprint{1502.03757}.

\bibitem[{\citenamefont{Cao et~al.}(2015)\citenamefont{Cao, Qin, and
  Bass}}]{Cao:2015hia}
\bibinfo{author}{\bibfnamefont{S.}~\bibnamefont{Cao}},
  \bibinfo{author}{\bibfnamefont{G.-Y.} \bibnamefont{Qin}}, \bibnamefont{and}
  \bibinfo{author}{\bibfnamefont{S.~A.} \bibnamefont{Bass}},
  \bibinfo{journal}{Phys. Rev. C} \textbf{\bibinfo{volume}{92}},
  \bibinfo{pages}{024907} (\bibinfo{year}{2015}), \eprint{1505.01413}.

\bibitem[{\citenamefont{Miller et~al.}(2007)\citenamefont{Miller, Reygers,
  Sanders, and Steinberg}}]{Miller:2007ri}
\bibinfo{author}{\bibfnamefont{M.~L.} \bibnamefont{Miller}},
  \bibinfo{author}{\bibfnamefont{K.}~\bibnamefont{Reygers}},
  \bibinfo{author}{\bibfnamefont{S.~J.} \bibnamefont{Sanders}},
  \bibnamefont{and}
  \bibinfo{author}{\bibfnamefont{P.}~\bibnamefont{Steinberg}},
  \bibinfo{journal}{Ann. Rev. Nucl. Part. Sci.} \textbf{\bibinfo{volume}{57}},
  \bibinfo{pages}{205} (\bibinfo{year}{2007}), \eprint{nucl-ex/0701025}.

\bibitem[{\citenamefont{Chen et~al.}(2022)\citenamefont{Chen, Jiang, Liu, Liu,
  and Zhao}}]{Chen:2021akx}
\bibinfo{author}{\bibfnamefont{B.}~\bibnamefont{Chen}},
  \bibinfo{author}{\bibfnamefont{L.}~\bibnamefont{Jiang}},
  \bibinfo{author}{\bibfnamefont{X.-H.} \bibnamefont{Liu}},
  \bibinfo{author}{\bibfnamefont{Y.}~\bibnamefont{Liu}}, \bibnamefont{and}
  \bibinfo{author}{\bibfnamefont{J.}~\bibnamefont{Zhao}},
  \bibinfo{journal}{Phys. Rev. C} \textbf{\bibinfo{volume}{105}},
  \bibinfo{pages}{054901} (\bibinfo{year}{2022}), \eprint{2107.00969}.

\bibitem[{\citenamefont{Song and Berrehrah}(2016)}]{Song:2016lfv}
\bibinfo{author}{\bibfnamefont{T.}~\bibnamefont{Song}} \bibnamefont{and}
  \bibinfo{author}{\bibfnamefont{H.}~\bibnamefont{Berrehrah}},
  \bibinfo{journal}{Phys. Rev. C} \textbf{\bibinfo{volume}{94}},
  \bibinfo{pages}{034901} (\bibinfo{year}{2016}), \eprint{1601.04449}.

\bibitem[{\citenamefont{Andronic et~al.}(2021)\citenamefont{Andronic,
  Braun-Munzinger, K\"ohler, Mazeliauskas, Redlich, Stachel, and
  Vislavicius}}]{Andronic:2021erx}
\bibinfo{author}{\bibfnamefont{A.}~\bibnamefont{Andronic}},
  \bibinfo{author}{\bibfnamefont{P.}~\bibnamefont{Braun-Munzinger}},
  \bibinfo{author}{\bibfnamefont{M.~K.} \bibnamefont{K\"ohler}},
  \bibinfo{author}{\bibfnamefont{A.}~\bibnamefont{Mazeliauskas}},
  \bibinfo{author}{\bibfnamefont{K.}~\bibnamefont{Redlich}},
  \bibinfo{author}{\bibfnamefont{J.}~\bibnamefont{Stachel}}, \bibnamefont{and}
  \bibinfo{author}{\bibfnamefont{V.}~\bibnamefont{Vislavicius}},
  \bibinfo{journal}{JHEP} \textbf{\bibinfo{volume}{07}}, \bibinfo{pages}{035}
  (\bibinfo{year}{2021}), \eprint{2104.12754}.

\bibitem[{\citenamefont{Group}(2020)}]{10.1093/ptep/ptaa104}
\bibinfo{author}{\bibfnamefont{P.~D.} \bibnamefont{Group}},
  \bibinfo{journal}{Progress of Theoretical and Experimental Physics}
  \textbf{\bibinfo{volume}{2020}} (\bibinfo{year}{2020}), ISSN
  \bibinfo{issn}{2050-3911}, \bibinfo{note}{083C01}.

\bibitem[{\citenamefont{Aad et~al.}(2013)}]{ATLAS:2013cia}
\bibinfo{author}{\bibfnamefont{G.}~\bibnamefont{Aad}} \bibnamefont{et~al.}
  (\bibinfo{collaboration}{ATLAS}), \bibinfo{journal}{JHEP}
  \textbf{\bibinfo{volume}{10}}, \bibinfo{pages}{042} (\bibinfo{year}{2013}),
  \eprint{1307.0126}.

\bibitem[{\citenamefont{Khachatryan et~al.}(2011)}]{CMS:2011oft}
\bibinfo{author}{\bibfnamefont{V.}~\bibnamefont{Khachatryan}}
  \bibnamefont{et~al.} (\bibinfo{collaboration}{CMS}), \bibinfo{journal}{Phys.
  Rev. Lett.} \textbf{\bibinfo{volume}{106}}, \bibinfo{pages}{112001}
  (\bibinfo{year}{2011}), \eprint{1101.0131}.

\bibitem[{\citenamefont{Acharya et~al.}(2022)}]{ALICE:2021edd}
\bibinfo{author}{\bibfnamefont{S.}~\bibnamefont{Acharya}} \bibnamefont{et~al.}
  (\bibinfo{collaboration}{ALICE}), \bibinfo{journal}{JHEP}
  \textbf{\bibinfo{volume}{03}}, \bibinfo{pages}{190} (\bibinfo{year}{2022}),
  \eprint{2108.02523}.

\bibitem[{\citenamefont{Sirunyan et~al.}(2017)}]{CMS:2017exb}
\bibinfo{author}{\bibfnamefont{A.~M.} \bibnamefont{Sirunyan}}
  \bibnamefont{et~al.} (\bibinfo{collaboration}{CMS}), \bibinfo{journal}{Eur.
  Phys. J. C} \textbf{\bibinfo{volume}{77}}, \bibinfo{pages}{269}
  (\bibinfo{year}{2017}), \eprint{1702.01462}.

\bibitem[{\citenamefont{Aaboud et~al.}(2018{\natexlab{b}})}]{ATLAS:2017prf}
\bibinfo{author}{\bibfnamefont{M.}~\bibnamefont{Aaboud}} \bibnamefont{et~al.}
  (\bibinfo{collaboration}{ATLAS}), \bibinfo{journal}{Eur. Phys. J. C}
  \textbf{\bibinfo{volume}{78}}, \bibinfo{pages}{171}
  (\bibinfo{year}{2018}{\natexlab{b}}), \eprint{1709.03089}.

\bibitem[{\citenamefont{Schenke et~al.}(2011)\citenamefont{Schenke, Jeon, and
  Gale}}]{Schenke:2010rr}
\bibinfo{author}{\bibfnamefont{B.}~\bibnamefont{Schenke}},
  \bibinfo{author}{\bibfnamefont{S.}~\bibnamefont{Jeon}}, \bibnamefont{and}
  \bibinfo{author}{\bibfnamefont{C.}~\bibnamefont{Gale}},
  \bibinfo{journal}{Phys. Rev. Lett.} \textbf{\bibinfo{volume}{106}},
  \bibinfo{pages}{042301} (\bibinfo{year}{2011}), \eprint{1009.3244}.

\bibitem[{\citenamefont{Shen et~al.}(2016)\citenamefont{Shen, Qiu, Song,
  Bernhard, Bass, and Heinz}}]{Shen:2014vra}
\bibinfo{author}{\bibfnamefont{C.}~\bibnamefont{Shen}},
  \bibinfo{author}{\bibfnamefont{Z.}~\bibnamefont{Qiu}},
  \bibinfo{author}{\bibfnamefont{H.}~\bibnamefont{Song}},
  \bibinfo{author}{\bibfnamefont{J.}~\bibnamefont{Bernhard}},
  \bibinfo{author}{\bibfnamefont{S.}~\bibnamefont{Bass}}, \bibnamefont{and}
  \bibinfo{author}{\bibfnamefont{U.}~\bibnamefont{Heinz}},
  \bibinfo{journal}{Comput. Phys. Commun.} \textbf{\bibinfo{volume}{199}},
  \bibinfo{pages}{61} (\bibinfo{year}{2016}), \eprint{1409.8164}.

\bibitem[{\citenamefont{Pang et~al.}(2012)\citenamefont{Pang, Wang, and
  Wang}}]{Pang:2012he}
\bibinfo{author}{\bibfnamefont{L.}~\bibnamefont{Pang}},
  \bibinfo{author}{\bibfnamefont{Q.}~\bibnamefont{Wang}}, \bibnamefont{and}
  \bibinfo{author}{\bibfnamefont{X.-N.} \bibnamefont{Wang}},
  \bibinfo{journal}{Phys. Rev. C} \textbf{\bibinfo{volume}{86}},
  \bibinfo{pages}{024911} (\bibinfo{year}{2012}), \eprint{1205.5019}.

\bibitem[{\citenamefont{Sollfrank et~al.}(1997)\citenamefont{Sollfrank,
  Huovinen, Kataja, Ruuskanen, Prakash, and Venugopalan}}]{Sollfrank:1996hd}
\bibinfo{author}{\bibfnamefont{J.}~\bibnamefont{Sollfrank}},
  \bibinfo{author}{\bibfnamefont{P.}~\bibnamefont{Huovinen}},
  \bibinfo{author}{\bibfnamefont{M.}~\bibnamefont{Kataja}},
  \bibinfo{author}{\bibfnamefont{P.~V.} \bibnamefont{Ruuskanen}},
  \bibinfo{author}{\bibfnamefont{M.}~\bibnamefont{Prakash}}, \bibnamefont{and}
  \bibinfo{author}{\bibfnamefont{R.}~\bibnamefont{Venugopalan}},
  \bibinfo{journal}{Phys. Rev. C} \textbf{\bibinfo{volume}{55}},
  \bibinfo{pages}{392} (\bibinfo{year}{1997}), \eprint{nucl-th/9607029}.

\bibitem[{\citenamefont{ALICE}(2022)}]{ALICE:2022qm}
\bibinfo{author}{\bibnamefont{ALICE}},
  \bibinfo{journal}{https://alice-figure.web.cern.ch/node/22314}
  (\bibinfo{year}{2022}).

\bibitem[{\citenamefont{CMS}(2022)}]{CMS:2022gvy}
\bibinfo{author}{\bibnamefont{CMS}}, \bibinfo{journal}{CMS-PAS-HIN-21-008}
  (\bibinfo{year}{2022}).

\end{thebibliography}

\end{document}